\documentclass[onecolumn,showpacs,11pt]{revtex4}
\usepackage{graphicx}
\usepackage{dcolumn}
\usepackage{bm}
\begin{document}
\newcommand{\hs}{\hspace*{0.5cm}}
\newcommand{\vs}{\vspace*{0.5cm}}
\newcommand{\be}{\begin{equation}}
\newcommand{\ee}{\end{equation}}
\newcommand{\bea}{\begin{eqnarray}}
\newcommand{\eea}{\end{eqnarray}}
\newcommand{\ben}{\begin{enumerate}}
\newcommand{\een}{\end{enumerate}}
\newcommand{\bde}{\begin{widetext}}
\newcommand{\ede}{\end{widetext}}
\newcommand{\nn}{\nonumber}
\newcommand{\crn}{\nonumber \\}
\newcommand{\non}{\nonumber}
\newcommand{\noi}{\noindent}
\newcommand{\al}{\alpha}
\newcommand{\la}{\lambda}
\newcommand{\bet}{\beta}
\newcommand{\ga}{\gamma}
\newcommand{\va}{\varphi}
\newcommand{\om}{\omega}
\newcommand{\pa}{\partial}
\newcommand{\fr}{\frac}
\newcommand{\bc}{\begin{center}}
\newcommand{\ec}{\end{center}}
\newcommand{\Ga}{\Gamma}
\newcommand{\de}{\delta}
\newcommand{\De}{\Delta}
\newcommand{\ep}{\epsilon}
\newcommand{\varep}{\varepsilon}
\newcommand{\ka}{\kappa}
\newcommand{\La}{\Lambda}
\newcommand{\si}{\sigma}
\newcommand{\Si}{\Sigma}
\newcommand{\ta}{\tau}
\newcommand{\up}{\upsilon}
\newcommand{\Up}{\Upsilon}
\newcommand{\ze}{\zeta}
\newcommand{\ps}{\psi}
\newcommand{\Ps}{\Psi}
\newcommand{\ph}{\phi}
\newcommand{\vph}{\varphi}
\newcommand{\Ph}{\Phi}
\newcommand{\Om}{\Omega}

\title{Question of Peccei-Quinn symmetry and quark masses\\ in the economical 3-3-1 model}

\author{P. V. Dong}
\email {pvdong@iop.vast.ac.vn} \affiliation{Institute of Physics,
VAST, 10 Dao Tan, Ba Dinh, Hanoi, Vietnam}
\author{H. T. Hung}
\email{hthung@grad.iop.vast.ac.vn} \affiliation{Department  of
Physics, Hanoi University of Education II, Phuc Yen, Vinh Phuc, Vietnam}
\author{H. N. Long}
\email{hnlong@iop.vast.ac.vn} \affiliation{Institute of Physics,
VAST, 10 Dao Tan, Ba Dinh, Hanoi, Vietnam}

\date{\today}

\begin{abstract}
We show that there is an infinite number of $U(1)$ symmetries like Peccei-Quinn symmetry in the 3-3-1 model with minimal scalar sector---two scalar triplets. Moreover, all of them are completely broken due to the model's scalars by themselves (notice that these scalars as known have been often used to break the gauge symmetry and generating the masses for the model's particles). There is no any residual Peccei-Quinn symmetry. Because of the minimal scalar content there are some quarks that are massless at tree-level, but they can get consistent mass contributions at one-loop due to this fact. Interestingly, axions as associated with the mentioned $U(1)$s breaking (including Majoron due to lepton-charge breaking) are all gauged away because they are also the Goldstone bosons responsible for the gauge symmetry breaking as usual.      

\end{abstract}


\maketitle

\section{\label{intro}Introduction}

There are obvious evidences that we must go beyond the standard model. The leading questions of which perhaps include neutrino oscillation, natural origin of masses and particularly Higgs mechanism, hierarchy problem between weak and Planck scale, and matter-antimatter asymmetry in the universe \cite{pdg}. In this work we will, however, be interested in alternatives concerning flavor physics. Why are there just three families of fermions? How are the families related, and what is the nature of flavor mixings and mass hierarchies?

We know that in the standard model, the $SU(2)_L$ symmetry is ``safe'' since its anomaly $\mathrm{Tr}[\{T_i,T_j\}T_k]=0$ for every $SU(2)_L$ representation. Each family of the standard model is anomaly free due to this fact, and the number of families on this ground can be left arbitrarily. For the reasons as stated, if we particularly extend $SU(2)_L$ to $SU(N)_L$ ($N\geq 3$), the anomaly $\mathrm{Tr}[\{T_i,T_j\}T_k]\neq 0$ for complex $SU(N)_L$ representations \cite{lang}. With a respective extension of fermion representations, $SU(2)_L$ singlets to $SU(N)_L$ singlets and $SU(2)_L$ doublets to $SU(N)_L$ fundamental multiplets or antimultiplets, each family now depends on the anomaly. It is cancelled if the number of multiplets equals to that of antimultiplets because their anomaly contributions are in the same size but opposite \cite{lang}. Thus, the number of families is an integral multiple of fundamental color number (three) in order to suppress that anomaly over total fermion content \cite{anoma}. It is noteworthy that  the families are now constrained on the background of anomaly cancellation (this disappears in the standard model). In this scenario we must require extra quarks and/or leptons to complete the representations. It is shown that the number of quark families must be smaller than or equals to $[33/(2N)]$ ($=5,\ 4,\ \mbox{or}\  3$ for $N= 3,\ 4,\ 5$, respectively) to ensure QCD asymptotic freedom condition. We deduce the number of families equals to three, coinciding with the observations.   

The simplest choice is $N=3$. Therefore, the gauge symmetry has the form $SU(3)_C\otimes SU(3)_L\otimes U(1)_X$ (thus named 3-3-1), where the strong interaction sector remains while the electroweak one is enlarged \cite{331r,331m}.  A fermion content satisfying all the requirements is
\bea
&&\psi_{aL} =\left( \nu_{a},e_{a},N^c_{a}\right) _{L}^{T}\sim
\left( 1,3,-1/3\right),\hs e_{aR}\sim \left( 1,1,-1\right), \crn
&& Q_{1L} =\left( u_{1},d_{1},U\right) _{L}^{T}\sim \left(
3,3,1/3\right),\hs Q_{\alpha L} =\left( d_{\alpha },-u_{\alpha
},D_{\alpha }\right)_{L}^{T}\sim \left( 3,3^{\ast },0\right), \crn
&& u_{aR}, U_R \sim \left(3,1,2/3\right),\hs  d_{aR}, D_{\al R}
\sim (3,1,-1/3), \label{eq1}\eea where $\alpha =\{2,3\}$ and $a=\{1,\al\}$ are family
indices. The quantum numbers as given in parentheses are
respectively based on $\left(SU\left(3\right)_{C},SU\left(
3\right)_{L},U\left(1\right)_{X}\right)$ symmetries. The $U$ and $D$ are exotic
quarks, while $N_R$ are right-handed neutrinos. The model is thus named the 3-3-1 model with right-handed neutrinos \cite{331r}. If these exotic leptons are not introduced, i.e. instead the third components are now included ordinary right-handed charged leptons, we have the minimal 3-3-1 model \cite{331m}.  
                     
The 3-3-1 gauge symmetry is broken through two stages: $SU(3)_L \otimes U(1)_X\longrightarrow SU(2)_L\otimes U(1)_Y\longrightarrow U(1)_{em}$. They are obtained by scalar triplets. One of the weaknesses of the mentioned 3-3-1 models that reduces
their predictive possibility is a plenty or complication in the
scalar sectors. The attempts on this direction to realize simpler
scalar sectors have recently been made. The first one is the 3-3-1  model with
right-handed neutrinos and minimal scalar sector---two triplets, \bea \chi &=&\left( \chi^{0}_1,\chi ^{-}_2,\chi
_{3}^{0}\right) ^{T}\sim \left(1,3,-1/3\right), \crn\phi &=&\left(
\phi ^{+}_1,\phi_2 ^{0},\phi _{3}^{+}\right)^T \sim \left(1,3,
2/3\right), \label{eq2}\eea with VEV given by
\be
\langle\chi\rangle=\fr{1}{\sqrt{2}}\left(%
\begin{array}{c}
  u \\
  0 \\
  \om \\
\end{array}%
\right),\hs
 \langle\phi\rangle=\fr{1}{\sqrt{2}}\left(%
\begin{array}{c}
  0 \\
  v \\
  0 \\
\end{array}%
\right), \label{vevp}\ee called the economical 3-3-1 model
\cite{ecn331}. The VEV $\om$ is responsible for the first stage of gauge symmetry breaking, while $v,u$ are for the second stage. The minimal 3-3-1 model with minimal scalar sector of two triplets has also been proposed in  Ref.
\cite{ecn331m}. Notice that due to the restricted scalar contents, these models often contain tree-level massless quarks that require corrections. The latter model \cite{ecn331m} has provided masses for quarks via higher-dimensional effective interactions, whereas the former one has produced quark masses via quantum effects \cite{dhhl}.      

The strong-CP question in the 3-3-1 models \cite{331m,331r} was  studied previously in \cite{pal}. The idea is that in these models by themselves we can find out extra global $U(1)$ symmetries which are naturally chiral responsible for the Peccei-Quinn symmetry \cite{pqsym}. The chiral means that there is at least a quark state which its left-handed and right-handed components have different charges under the mentioned $U(1)$. Moreover, the $U(1)$ current at the quantum level has to be anomalously non-conserved as accommodated by the (chiral) color anomaly $[SU(3)_C]^2 U(1)$, which must be nonzero \cite{pqsym}. There is no extra scalar required unlike the case of the standard model \cite{pqsym,axion}. The models' scalars by themselves can be reserved for breaking the $U(1)$ under which the models are made free from the strong-CP problem. The nature of axion in these models, however, takes quite the same status as in the standard model extensions \cite{pqsym,axion} which has been extensively studied \cite{pal}.      

For the economical 3-3-1 model (which is being investigated in this work) \cite{ecn331}, however, a recent paper \cite{natur} has argued that due to an $U\left( 1\right) _{PQ}$ global symmetry like the Peccei-Quinn symmetry \cite{pqsym}, there exist some quarks that are massless up to any order of the perturbative theory. Consequently, the economical 3-3-1 model as constructed in \cite{ecn331} is not correct. The authors stated that this is due to a residual symmetry $U(1)''_{PQ}$ of $U(1)_{PQ}$ surviving after the symmetry breaking that prevents the quarks from getting mass. 

In this paper, we reconsider  in details the above puzzle and show that after spontaneous symmetry breaking, the residual symmetries in the
economical 3-3-1 model including $U(1)_{PQ}$ above are \emph{only} the $U(1)_{em}$ (and $U(1)_B$ of baryon number if it was included). There is no such $U(1)''_{PQ}$ at all. Indeed, all of the $U\left( 1\right) _{PQ}$ symmetries are totally broken as the lepton number. Therefore there is no reason why the quark masses cannot be generated. All quarks can get masses due to the scalar vacuum values that break this symmetry. As examples, the complete loop corrections as well as the higher-dimensional effective interactions \cite{wbg} responsible for quark masses are evaluated in details to prove our judgment. We will also show that in this model there are not axion and Majoron associated the broken symmetries, which are unlike other extensions \cite{pqsym,axion,majoron} of the standard model. They are already gauged away since they are just the Goldstone bosons responsible for the gauge symmetry breaking.  

The rest of this work is follows. In Sec. \ref{natural}, we stress on the model. Possible Peccei-Quinn like symmetries will be obtained. The
breakdown of all these $U(1)$s as well as the gauge symmetry will be examined. In Sec. \ref{oneloop} we will give a complete evaluation of all the one-loop mass corrections as well as five-dimensional effective mass operators responsible for the massless quarks. Finally we make
conclusions in the last section--Sec. \ref{con}.

\section{\label{natural} The model and Peccei-Quinn like symmetries}

The gauge symmetry of the model is $SU(3)_C\otimes SU(3)_L\otimes U(1)_X$. The particle content is defined in equations (\ref{eq1},\ref{eq2}).  The electric charge operator is given by
\begin{equation}
Q=T_3- \fr{1}{\sqrt{3}}T_8+X,\label{ecqddd}
\end{equation} where $T_i$ $(i=1,2,3,...,8)$ and $X$ are the
charges of $SU(3)_L$ and $U(1)_X$, respectively. The standard model hypercharge operator is thus identified as $Y=-(1/\sqrt{3}) T_8+ X$.
This model does not contain exotic electric charges, i.e. the
exotic quarks have electric charges like ordinary quarks:
$Q\left(D\right) =-1/3$ and $Q\left( U \right) =2/3$.

The most general Yukawa interactions are given by\bea {\mathcal
L}_{\mathrm{Y}}&=&h^e_{ab}\overline{\psi}_{aL}\phi
e_{bR}+h^\nu_{ab}\ep_{mnp}(\overline{\psi}^c_{aL})_m(\psi_{bL})_n(\phi)_p
\crn && +h^U\overline{Q}_{1L}\chi
U_{R}+h^D_{\al\beta}\overline{Q}_{\al L}\chi^* D_{\beta R}+h^d_{a}\overline{Q}_{1 L}\phi d_{a R}+h^u_{\al
a}\overline{Q}_{\al L}\phi^* u_{aR}\crn
&&+s^u_{a}\overline{Q}_{1L}\chi u_{aR}+s^d_{\al
a}\overline{Q}_{\al L}\chi^* d_{a R} +s^D_{
\al}\overline{Q}_{1L}\phi D_{\al R}+s^U_{\al }\overline{Q}_{\al
L}\phi^* U_{R}\crn
&&+ H.c.,\label{y2}\eea where $m$, $n$ and $p$ stand
for $SU(3)_L$ indices. In \cite{dhhl} we have shown that
at the tree level one up-quark and two down-quarks are massless.
However, the one-loop corrections can give them consistent masses.
In this work we will revisit those quantum effects by giving a
complete calculation when including a realistic mixing of all the
three families of quarks as well. We are thus showing that the
results in \cite{natur} which contrast with ours are not correct.

As the lepton triplets stand, the lepton number in this model does not commute with the
gauge symmetry. In fact, it is a residual symmetry of a new-lepton
charge $\mathcal{L}$ given by \cite{changlong} \be
L=\fr{4}{\sqrt{3}}T_8+\mathcal{L}.\ee The $\mathcal{L}$ charges of
the model multiplets can be obtained as \be
\mathcal{L}(\psi_{aL},Q_{1L}, Q_{\al
L},\phi,\chi,e_{aR},u_{aR},d_{aR},U_R,D_{\al R})=\fr 1 3,-\fr 2
3,\fr 2 3,-\fr 2 3,\fr 4 3,1,0,0,-2,2,\ee respectively. Also, it
is easily checked that $L(U)=-L(D)=L(\phi_3)=-L(\chi_{1,2})=-2$.
All the other quarks and scalars have zero lepton-number, $L=0$.
It is worth emphasizing that the residual $L$ is {\it
spontaneously} broken by $u$ due to $L(\chi^0_1)=2$, which is
unlike the case of the standard model. Notice that the Yukawa couplings $s$'s
violate $\mathcal{L}$, while the $h$'s do not.

Following \cite{natur}, we introduce a global $U(1)_{H}$ symmetry in addition to the gauge symmetry, i.e. 
\be SU(3)_C\otimes SU(3)_L\otimes U(1)_X\otimes U(1)_{H}.\ee The condition for this symmetry playing the role like Peccei-Quinn (handedness or chiral) symmetry is the chiral anomaly $[SU(3)_C]^2 U(1)_{H}\neq 0$ as mentioned. On the other hand, since the Yukawa interactions invariant under this symmetry, the relations on the charges of any $U(1)_{H}$ group are \bea &&-H_{Q_{1}}+H_{U}+H_{\chi } =0,
\qquad
-H_{Q}+H_{D}-H_{\chi }=0,  \label{e1} \\
&&-H_{Q_{1}}+H_{u}+H_{\chi } =0,\qquad
-H_{Q}+H_{d}-H_{\chi
}=0, \\
&&-H_{Q_{1}}+H_{d}+H_{\phi } =0,\qquad
-H_{Q}+H_{u}-H_{\phi
}=0, \\
&&-H_{Q_{1}}+H_{D}+H_{\phi } =0,\qquad
-H_{Q}+H_{U}-H_{\phi } =0, \\
&&-H_{\psi }+H_{e}+H_{\phi} =0,\qquad\hs 2H_{\psi }+H_{\phi
}=0, \label{e5} \eea where the notation $H_{\Psi }$ means as the
$U\left( 1\right)_H $ charge of the $\Psi $ multiplet (remind that $H$ is similar to $X$ in \cite{natur}). Notice that  all the other parts of the Lagrangian are obviously conserved under this symmetry. Using the relations, the chiral anomaly is rewritten as \be [SU(3)_C]^2U(1)_{H}\sim 2H_\chi+H_\phi\neq 0.\label{ano}\ee 

In solving equations (\ref{e1}-\ref{e5},\ref{ano}), we also denote $H$ as a collection of partial solutions $H_{\Psi}$ in order, and having remarks as follows
\ben \item The solution is scale invariance, i.e. if $H$ is solution, then $cH$ ($c\neq 0$) does. 
\item Two solutions called to be different (i.e. linearly independent) if they are not related by scale invariance transformations.
\item The solutions that contain linearly-independent subsolutions, e.g. $(H_\phi,H_\chi)=(0,1),\ (1,0)$, or $(1,1)$, respectively, are different.    
\item The different solutions will define different Peccei-Quinn like symmetries, respectively. So, the number of different solutions found obeys just that of Peccei-Quinn like symmetries presenting in our model.  
\een

The charge relations (\ref{e1}-\ref{e5}) yield degenerate equations. Indeed, they can equivalently be rewritten via seven independent equations as follows 
\bea H_{u}&=&H_{U},\hs H_{d}=H_{D},\hs H_{\psi}=-H_{\phi}/2,\hs H_{e}=-3H_{\phi}/2,\crn
H_{u}-H_{d}&=&H_{\phi}-H_{\chi},\hs H_{u}-H_Q=H_{\phi},\hs H_{u}+H_{d}=H_Q+H_{Q_1}. \eea We have 10 variables, while there are 7 equations. Hence, there is an infinite number of solutions (certainly satisfying (\ref{ano}) too). For instance, put $H_\phi=0$. We have $H_\psi=H_e=0$, $H_u=H_U=H_Q$, $H_d=H_D=H_{Q_1}$, and $H_u-H_d=-H_\chi$. The solutions of this kind are thus given dependently on two parameters $a\equiv H_\chi \neq 0$ and $b\equiv H_u$ such as
\be H(\phi,\chi,\psi,e,u,U,Q,d,D,Q_1)=(0,a,0,0,b,b,b,a+b,a+b,a+b).\ee Since $a,b$ are arbitrary, there are an infinity of different solutions corresponding to whatever pairs $(a,b)$ are linearly independent, for example, $(a,b)=(0,1),\ (1,0),\ (1,1), (1,2)$ and so on.

In Table \ref{table 1}, we list three of Peccei-Quinn like symmetries in which the first one (second line) is given in \cite{natur} that was solely claimed and marked as $U(1)_{PQ}$.
\begin{table}[th]
\caption{Three chiral symmetries taken as examples in the economical 3-3-1
model.} \label{table 1}\centering
\begin{tabular}{cccccccc}\hline  $Q_{\alpha L}$ & \,\, $Q_{1L}$ & \,\,
($u_{aR}$, $U_{R}$) & \,\,($d_{aR}$, $D_{\alpha R}$) & \,\,
$\psi_{aL}$ & \,\, $ e_{aR} $ & \,\, $\phi $ & \, $\chi $
\\ \hline  $-1$ & $1$ & $0$ & $0$ & $-1/2$
& $-3/2$ & $1$ & $1$ \\ \hline
  $1$ & $2$ & $1$ & $2$ & $0$ & $0$ & $0$ & $1$
\\ \hline
  $1$ & $2$ & $2$ & $1$ & $-1/2$ & $-3/2$ & $1$ & $0$ \\ \hline
\end{tabular}
\end{table}

The symmetries $SU(3)_L\otimes U(1)_X\otimes U(1)_H$ [where the first two are local (gauged) while the last one is global; all of them appear as group factors in the direct products] are broken by $\chi$ and $\phi$ (when they get VEVs) since these scalars are charged under all the groups. Because every scalar carries multi-charges, $SU(3)_L$, $U(1)_X$, and $U(1)_H$, it will break all these groups when it develops VEV.  In general (which it will happen in this model) we can have one Goldstone boson while more than one of the generators associated are broken. The important thing is to account the generators of the unbroken residual symmetry.  Any generator of the residual symmetry has the form $\al_i T_i + \gamma X +\delta H$. If $\delta=0$ (this will also happen in the presenting model), $U(1)_H$ is decoupled and totally broken. The Goldstone boson associated with $U(1)_H$ firstly seem to be physical since this group is global. However, it also carries the other broken charges of the $SU(3)_L\otimes U(1)_X$ gauge symmetry which it will be finally gauged away. In this case, the anomalies associated with $U(1)_H$ are no concerned except the one responsible for a Peccei-Quinn like symmetry as given above. If $\delta \neq 0$ which is not the case in the presenting model, $U(1)_H$ will be also gauged since $T_i$ and $X$ in the combination are actually gauged charges. Any anomaly associated with $U(1)_H$, e.g. $[SU(3)_L]^2U(1)_H$, must be taken into account.        

Fortunately, there is no residual symmetry associated with the $U(1)_H$ above after the spontaneous symmetry breaking which contradicts with \cite{natur}. Prove: suppose that there is such one, denoted by $U(1)_{PQ}$ (let us note that in \cite{natur} they called it as $U(1)''_{PQ}$ instead). Since it is survival and conserved after the spontaneous symmetry breaking as supposed, it has the form as a combination of diagonal generators \be PQ=\al T_3+\beta T_8 +\gamma X + \delta H\hs\hs (\delta \neq 0).\ee Also, the charge $PQ$ has to annihilate the vacuums, \be PQ(\langle \phi\rangle)=0,\hs\hs PQ(\langle \chi\rangle)=0.\label{huyvev}\ee All these are quite similar  as in obtaining the electric charge operator responsible for electric charge conservation after the electroweak symmetry breaking. We therefore have equations:
\bea
\fr{\al}{2} +\fr{\beta}{2\sqrt{3}}+\ga X_\chi + \delta H_\chi &=& 0,\crn
-\fr{\al}{2} +\fr{\beta}{2\sqrt{3}}+\ga X_\phi + \delta H_\phi &=& 0,\crn
-\fr{\beta}{\sqrt{3}}+\ga X_\chi + \delta H_\chi &=& 0.\nn
\eea
Combining all three equations, we deduce $2H_{\chi}+H_{\phi}=0$ that contradicts to (\ref{ano}). Therefore, there is no residual symmetry of $U(1)_H$. All the Peccei-Quinn like $U(1)_H$ symmetries are completely broken along with the gauge symmetry breaking. (We can extend this prove for a supposed non-Abelian residual symmetry with generators $\al_i T_i + \gamma X +\delta H$; the conclusion will be unchanged.)

In addition if (\ref{ano}) is satisfied and leaving $\delta$ arbitrary, we have $\delta=0$ and $\beta=-\al/\sqrt{3}$ by solving the above three equations. With the note of the $X$-charge values of the scalars as given in the introduction (that embed the electric charge operator in the form as defined by  (\ref{ecqddd})), we get also $\gamma=\al$. Therefore we have $PQ=\al Q$ as a solution to finding the electric charge operation (that certainly contradicts to (\ref{ano}) since $Q$ is vectorlike). If one includes baryon number $B$ as well (since $B_\phi=B_\chi=0$), it results \be PQ=\al Q + \xi B. \ee Only vectorlike symmetries (i.e. non Peccei-Quinn) such as $X$, $B$ might have surviving residual symmetries after the spontaneous symmetry breaking by the model's scalars. To conclude, with $SU(3)_L\otimes U(1)_X\otimes U(1)_H$ symmetries, we have $8+1+1=10$ generators. Only one, $Q$, is conserved by VEVs. Thus there are 9 broken generators. 

The scalar potential has the form \cite{dls1}
\bea V(\chi,\phi) &=& \mu_1^2 \chi^\dag \chi + \mu_2^2
\phi^\dag \phi + \la_1 ( \chi^\dag \chi)^2 + \la_2 ( \phi^\dag
\phi)^2\crn &  & + \la_3 ( \chi^\dag \chi)( \phi^\dag \phi) +
\la_4 ( \chi^\dag \phi)( \phi^\dag \chi). \label{poten} \eea             
With the two scalar triplets, we have totally twelve real scalar components. Eight of them are Goldstone bosons (massless) completely eaten by corresponding eight weak gauge bosons $W^\pm,\ Z$ and new $Z',\ X^{0,0*},\ Y^{\pm}$ of the $[SU(3)_L\otimes U(1)_X]/U(1)_{em}$ quotient group. The rest is Higgs particles (massive): two neutral $h,\ H$ and two charged $H^{\pm}$ (where $h$ is the standard model like Higgs particle). For details, see \cite{dls1}. Here we give a summary of the scalar content:
 \be \phi=\left(%
\begin{array}{c}
  G^+_{W} \\
  \fr{1}{\sqrt{2}}(v + h +  iG_Z) \\
  H^+ \\
\end{array}%
\right),\hs \hs \chi=\left(%
\begin{array}{c}
 \fr{1}{\sqrt{2}}u +  G^0_{X}\\
   G^-_{Y} \\
  \fr{1}{\sqrt{2}}(\om + H + i G_{Z'}) \\
\end{array}%
\right).\label{potenn35aa} \ee
There are 9 broken generators as mentioned while there are only 8 Goldstone bosons. At least one Goldstone boson is the so-called composite particle as the one responsible for both broken $H$ and some electroweak gauge generator. There is no axion. Also, there is no Majoron (note that lepton number is broken by $\chi^0_1$ since it carrys this charge). These particles unlike those which have been actually studied in other extensions are unphysical since they can dynamically been removed by the electroweak (local) gauge symmetry with an appropriate gauge transformation associated with their gauge charges \cite{pon11}.  

To be concrete, in the next section we will generate masses for all fermions. However, before doing that it is worth to mention on the spontaneous $U(1)_H$ symmetry breaking via the scalar potential with stressing on the scalar field self-interactions and mass-terms after shifting the scalar fields by their VEVs \cite{dls1}:
\bea 
V_{mass}&=& \la_1 (uS_1+\om S_3)^2+\la_2 v^2
S^2_2 +\la_3 v (uS_1+\om S_3)S_2\crn
&&+\fr{\la_4}{2}(u\phi^+_1+v\chi^+_2+\om
\phi^+_3)(u\phi^-_1+v\chi^-_2+\om \phi^-_3),
\label{pote11}\\
V_{int} &= &\la_1
(\chi^\dagger\chi)^2+\la_2(\phi^\dagger\phi)^2+\la_3
(\chi^\dagger\chi)(\phi^\dagger\phi)+\la_4 (\chi^\dagger\phi)(\phi^\dagger\chi)\crn && +
(2\la_1uS_1+\la_3 v
S_2+2\la_1\om S_3)(\chi^\dagger\chi)+(\la_3uS_1+2\la_2 v S_2+\la_3\om S_3)(\phi^\dagger\phi) \crn
&&+\fr{\la_4}{\sqrt{2}}(u\phi^-_1+v\chi^-_2+\om
\phi^-_3)(\chi^\dagger\phi)+H.c.,
\label{potenn4} \eea which can be obtained with the replacement of $\phi \rightarrow \phi + \langle \phi\rangle$ and $\chi\rightarrow \chi + \langle\chi\rangle$, where the VEVs are given in (\ref{vevp}) and $\chi,\phi$ are now physical fields with vanishing VEVs. (For details, see \cite{dls1}.) The neutral scalar fields have been denoted as \be \chi^0_1  =  \fr{S_1 +i
A_1}{\sqrt{2}},\hs \chi^0_3  =  \fr{S_3 + i A_3}{\sqrt{2}},\hs
\phi^0_2 = \fr{S_2 + i A_2}{\sqrt{2}}.\label{potenn5}\ee 

Since $2H_\chi + H_\phi \neq 0$ at least one of $H_\chi,H_\phi$ is nonzero. Therefore, the $U(1)_H$ symmetry is spontaneously broken by $\chi$ or $\phi$. In (\ref{pote11}) there must be some scalar masses obtained by breaking $U(1)_H$ at least one time when two among four scalars in original quartic interactions (\ref{poten}) getting the VEVs. After the spontaneous symmetry breaking, the second and third lines of $V_{int}$ that consist of triple scalar interactions possess terms violating this symmetry explicitly, due to $\chi$ or $\phi$ having $H$-charge. Here the first line of $V_{int}$ contains quartic scalar interactions that always conserve $U(1)_H$.  Notice that the $U(1)_H$ symmetry is also broken spontaneously in some Yukawa interactions  responsible for fermion masses after the scalars getting the VEVs. This is the first time we observe collective symmetry breaking phenomena (spontaneously or explicitly) due to the interplay of $\chi$ and $\phi$. All those will be responsive for generating realistic quark masses via quantum effects or higher-dimensional effective operators \cite{wbg} as shown below.  

We know that from the potential minimization conditions \cite{dls1}, the VEV of $\chi$ is generally given as $\langle \chi \rangle = \fr{1}{\sqrt{2}}(u,0,\om)$, where $u$ can in principle be nonzero so that the model is consistent as it  has often been taken \cite{ecn331}. In addition, this VEV arrangement can be related to that in the ordinary 3-3-1 model such as $\fr{1}{\sqrt{2}}(0,0,\om')$ \cite{331r} by a gauge symmetry transformation, for example, \be (u,0,\om)^T=e^{i\fr{u'}{\om'}\la_5}(0,0,\om')^T,\ee where 
\bea e^{i\fr{u'}{\om'}\la_5}=\left(\begin{array}{ccc} \cos \fr{u'}{\om'} & 0 & \sin \fr{u'}{\om'}\\
0& 1 & 0 \\
-\sin \fr{u'}{\om'} & 0 & \cos \fr{u'}{\om'}\end{array}\right),\eea 
$u\equiv \om'\sin \fr{u'}{\om'}=u'-\fr{1}{6}\fr{u'^3}{\om'^2}+\cdots\simeq u'$ and $\om\equiv \om'\cos \fr{u'}{\om'}=\om'-\fr 1 2 \fr{u'^2}{\om'}+\cdots \simeq \om'$ with $u'\ll \om'$ (equivalently, $u\ll \om$ as expected \cite{ecn331}). A question therefore arising is if the consequences and results of the economical 3-3-1 model can be derived from those of the ordinary 3-3-1 model, and even our conclusions as well as the model under consideration are not new? It is not correct because the Lagrangian is invariant under the gauge transformation, thus all the results as given retain. There are the consequences of the economical 3-3-1 model such as the mixing between the standard model gauge bosons and the new gauge bosons, $W-Y$ and $Z-W_4$, and mixing between ordinary quarks and exotic quarks \cite{ecn331,dhhl} that do not appear in the ordinary 3-3-1 model \cite{331r}. 

For details, in the new basis by the above gauge transformation, all the fields such as fermions, gauge bosons, scalars must be shifted correspondingly. For example, the fermion triplets/antitriplets are transformed by $e^{-i\fr{u'}{\om'}\la_5}$, which become now $(\nu_L-\fr{u}{\om} N^c_R, e_L, N^c_R+\fr{u}{\om} \nu_L)$, $(u_{1 L} -\fr{u}{\om} U_{L}, d_{1L}, U_{L}+\fr{u}{\om}u_{1 L})$ and $(d_{\al L} -\fr{u}{\om} D_{\al L}, -u_{\al L}, D_{\al L}+\fr{u}{\om}d_{\al L})$ where we have taken $u'\simeq u$, $\om'\simeq \om$ and $u\ll \om$. Notice that the right-handed fermion singlets remain unchanged. The Yukawa interactions are given similarly as above. It is easily checked that in this case the mixing between ordinary quarks and exotic quarks arises from the couplings of the third components $U_{L}+\fr{u}{\om}u_{1 L}$ and $D_{\al L}+\fr{u}{\om}d_{\al L}$ to the right-handed quarks $u_R,U_R$ and $d_R,D_R$, respectively via the VEV $\om'$ (thus $\om$). Similarly we can obtain that for the gauge bosons. Therefore, all the mass Lagrangians, mass eigenvalues, and mixings in the fermion and gauge boson sectors are given as before \cite{dhhl,ecn331} independent of the gauge transformations. In this model we have non-normal interactions for charged current and neutral current \cite{ecn331} that disappear in the ordinary 3-3-1 model. The consequences of the 3-3-1 model with right-handed neutrinos when we consider $u\neq 0$ or $u=0$ are very different.                       
Only if $u=0$ from the beginning, the model's results are a consequence of the known one \cite{331r}.  

Even if we work in the new basis by the above gauge transformation, the first condition presented below Eq. (\ref{huyvev}) still remains. Indeed, we were working in a basis $(u\neq 0)$ where the Gell-Mann matrices take the standard forms. In the new basis, the $PQ$ becomes 
\bea &&PQ' = e^{-i\fr{u'}{\om'}\la_5}PQ e^{i\fr{u'}{\om'}\la_5}\crn
&&=\left(\begin{array}{ccc} \left(\fr{\al}{2}
+\fr{\beta}{2\sqrt{3}}+\ga X +\de H \right)\cos\fr{u'}{\om'}(2-\cos \fr{u'}{\om'})   & & \\
+\left(-\fr{\beta}{\sqrt{3}}+\ga X +\de H\right)\sin^2\fr{u'}{\om'} &0  & \left(\fr{\al}{2}+\fr{\sqrt{3}\beta}{2}\right)\cos\fr{u'}{\om'}\sin\fr{u'}{\om'} \\ \\ \\
0 & -\fr{\al}{2}
+\fr{\beta}{2\sqrt{3}} & 0\\
 &+\ga X +\de H  & \\ \\
\left(\fr{\al}{2}+\fr{\sqrt{3}\beta}{2}\right)(2-\cos\fr{u'}{\om'})\sin\fr{u'}{\om'} & 0  &  \left(\fr{\al}{2}
+\fr{\beta}{2\sqrt{3}}+\ga X +\de H \right)\sin^2\fr{u'}{\om'}\\
 &  & +\left(-\fr{\beta}{\sqrt{3}}+\ga X +\de H\right)\cos\fr{u'}{\om'}(2-\cos\fr{u'}{\om'})\end{array}\right).\nn\eea The $PQ'$ has to annihilate $\langle \chi'\rangle = \fr{1}{\sqrt{2}}(0,0,\om')^T$. We have 
 \bea 
 \fr{\al}{2}+\fr{\sqrt{3}\beta}{2}&=&0,\label{tudong}\\
 \left(\fr{\al}{2}
+\fr{\beta}{2\sqrt{3}}+\ga X_\chi +\de H_\chi \right)\sin^2\fr{u'}{\om'} +\left(-\fr{\beta}{\sqrt{3}}+\ga X_\chi +\de H_\chi \right)\cos\fr{u'}{\om'}(2-\cos\fr{u'}{\om'})&=&0.\eea The second equation must be satisfied with any $u',\om'$. We deduce the first and the last equations as presented below (\ref{huyvev}). Note that the (\ref{tudong}) is also satisfied since it is a consequence of these two conditions when we let the first equation be subtracted by the last one.  

Finally, let us remark on the hierarchy $u\ll\om$ within the selected vacuum $(u,0,\om)$, even stronger $u\ll v \simeq 246\ \mathrm{GeV}\ll \om$, against the quantum correction. (These inequalities have been previously determined in \cite{ecn331} from the physical view points due to the model's consistency.) It is noted that the $\chi^0_1$ carries a lepton charge with two units. Also, if the lepton number is an exact symmetry, but not spontaneously broken, its VEV vanishes, i.e. $u=0$. When this field develops a VEV, $u\neq 0$, it will break the lepton symmetry spontaneously. We would expect that the lepton symmetry will prevent the $u$ naturally small as actually obtained in other models \cite{dongdongdong}. On the other hand, we do not have any physical Higgs field associated with the scale $u$ since $G_X$ is already a Goldstone boson, as can be checked from (\ref{potenn35aa}). The quantum correction to the mass of $G_X$ should vanish as a consequence of the gauge symmetry. The stabilization of the scale $u$ is therefore followed. A detailed study on this subject is interesting but out of the scope of this article. It should be published elsewhere.          

\section{\label{oneloop}Fermion masses}

In this model, the masses of charged leptons are given at the tree level as usual while the neutrinos can get consistent masses at the one-loop level as explicitly pointed out in Ref. \cite{dls2}. The implication of the higher-dimensional effective operators responsible for the neutrino masses has also been given therein. 

Let us now concentrate on masses of quarks that can be divided into two sectors: up type quarks $(u_a,U)$ with electric charge
$2/3$ and down type quarks $(d_a,D_\al)$ with electric charge $-1/3$. From (\ref{y2}) and (\ref{vevp}) we can obtain the mass matrix of the up type quarks $(u_1,
u_2, u_3, U)$: \be
M_{\mathrm{up}} = \fr{1}{\sqrt{2}}\left(%
\begin{array}{cccc}
  -s^u_{1} u & -s^u_{2}u  & -s^u_{3}u & -h^U u \\
 h^u_{2 1} v  & h^u_{2 2} v & h^u_{2 3} v & s^U_{2} v \\
 h^u_{3 1} v  & h^u_{3 2} v & h^u_{3 3} v & s^U_{3} v  \\
  -s^u_{1} \om  & -s^u_{2} \om & -s^u_{3} \om & -h^U \om \\
\end{array}%
\right), \label{upqmasstu} \ee and the mass matrix of down type quarks $(d_1, d_2, d_3, D_2,
D_3)$: \be M_{\mathrm{down}} = - \fr{1}{\sqrt{2}}
\left(%
\begin{array}{ccccc}
  h^d_{1} v  &  h^d_{2} v  &  h^d_{3} v  &  s^D_{2} v  & s^D_{3} v \\
  s^d_{21} u &  s^d_{22} u &  s^d_{23} u &  h^D_{22} u &  h^D_{23} u \\
   s^d_{31} u &  s^d_{32} u &  s^d_{33} u  & h^D_{32} u  & h^D_{33} u  \\
   s^d_{21} \om & s^d_{22} \om & s^d_{23} \om & h^D_{22} \om & h^D_{23} \om  \\
  s^d_{31} \om & s^d_{32} \om & s^d_{33} \om & h^D_{32} \om & h^D_{33}\om \\
\end{array}%
\right).
 \label{upqmasstdow2} \ee The first and last rows of (\ref{upqmasstu}) are proportional. Similarly, the second and fourth rows of
(\ref{upqmasstdow2}) are proportional, while the third and
last rows of this matrix take the same situation. Hence, in this model the tree level quark spectrum contains three massless
eigenstates (one up and two down quarks). So, what are the causes? 

There are just two: first all these degeneracies are due to the $\chi$ scalar only (with the presence of VEVs $u,\om$), {\it not} $\phi$; second, the Yukawa couplings of the first and third component of quark triplets/antitriplets to right-handed quarks in those degenerate rows are the same due to $SU(3)_L$ invariance. Obviously, the vanishing quark masses are not a consequence of the $U(1)_H$ symmetry because it actually happens even if we choose $H_\chi =0$ (in this case the Peccei-Quinn like symmetry resulting from only $H_\phi \neq 0$ does not give any constraint on the massless quark sector). At the one-loop level, all the degeneracies will be separated due to contribution of $\phi$ as well (see Appendix B of \cite{dhhl}). In such case, the one-loop mass corrections also collectively break the $U(1)_H$ symmetry since both the scalars $\chi,\phi$ are being taken into account, i.e. for those relevant quarks $2H_\chi + H_\phi$ are always nonzero at the one loop level.       

In \cite{dhhl}, we have already shown that all the tree level massless quarks can get consistent masses at the one-loop level. There, the light quarks and/or mixings of light quarks with exotic quarks got mass contributions. The exotic quark masses are reasonably large and took as a cutoff scale, thus no correction is needed. So why the recalculations as given in Ref. \cite{natur} for the quark masses, that consequently contradict to ours, are incomplete? This is due to the fact that they included even mass corrections for heavy exotic quarks as well. In this case, the cutoff scale of the theory must be larger than the exotic quark masses. As a result, under this cutoff scale all the physics is sensitive. There must be contributions coming from the $\phi$ scalar as well as ordinary active quarks where the flavor mixing must present. Let us remind that Ref. \cite{natur} in this case accounts for the $\chi$ contribution only. Thus the masslessness would remain as a result of two points mentioned above.  

The above analysis also means that all quarks will get masses if both $\chi$ and $\phi$ contribute so that $2 H_\chi + H_\phi \neq 0$ to ensure (\ref{ano}). This can explicitly be understood via an analysis of the effective mass operators \cite{wbg} responsible for quarks below.

\subsection{One-loop corrections} 

The analysis given below is for the up type quark sector only. That for the down type quark sector can be done similarly and got the same conclusion as the up type quarks. After the one-loop corrections, the mass matrix (\ref{upqmasstu}) looks like
\be M_{\mathrm{up}} = \fr{1}{\sqrt{2}}\left(%
\begin{array}{cccc}
  -s^u_{1} u +\Delta_{11}& -s^u_{2}u +\Delta_{12} & -s^u_{3}u +\Delta_{13}& -h^U u +\Delta_{14}\\
 h^u_{2 1} v  +\Delta_{21}& h^u_{2 2} v +\Delta_{22}& h^u_{2 3} v +\Delta_{23}& s^U_{2} v +\Delta_{24}\\
 h^u_{3 1} v  +\Delta_{31}& h^u_{3 2} v +\Delta_{32}& h^u_{3 3} v +\Delta_{33}& s^U_{3} v +\Delta_{34} \\
  -s^u_{1} \om  +\Delta_{41}& -s^u_{2} \om +\Delta_{42}& -s^u_{3} \om +\Delta_{43}& -h^U \om +\Delta_{44}\\
\end{array}%
\right), \label{upcor} \ee where $\Delta_{ij}$ are all possible one-loop corrections. Obviously this matrix gives all nonzero masses if the first and last rows are not in proportion. To show that tree-level degeneracy separated (i.e. these two rows are now not proportional) it is only necessary to prove the following submatrix:
\be M_{uU} = \fr{1}{\sqrt{2}}\left(%
\begin{array}{cc}
  -s^u_{1} u +\Delta_{11}& -h^U u +\Delta_{14}\\
   -s^u_{1} \om  +\Delta_{41}& -h^U \om +\Delta_{44}\\
\end{array}%
\right) \label{upcor1} \ee   
having nonzero determinant with general Yukawa couplings and VEVs. Two conditions below should be clarified: (i) The tree-level properties as implemented by the two points above must be broken, 
\bea     &&\left\{   \begin{array}{c} 
      \frac{\Delta_{11}}{\Delta_{14}}\neq  \frac{s^u_1}{h^U} \\
       \frac{\Delta_{41}}{\Delta_{44}}\neq \frac{s^u_1}{h^U} \\
   \end{array}\right.,\label{29dd}\\
   &&\left\{   \begin{array}{c} 
      \frac{\Delta_{11}}{\Delta_{41}}\neq  \frac{u}{\om} \\
       \frac{\Delta_{14}}{\Delta_{44}}\neq \frac{u}{\om} \\
   \end{array}\right..\label{ddd}
   \eea Since, by contrast if one of these systems is unsatisfied, which is the case as analyzed in \cite{natur}, one quark remains massless. (ii) The matrix (\ref{upcor1}) has nonzero determinant:
   \bea \mathrm{det} M_{uU} &=&\fr 1 2 \left[s^u_1(\om \Delta_{14}-u\Delta_{44})+h^U(u\Delta_{41}-\om \Delta_{11})+\Delta_{11}\Delta_{44}-\Delta_{41}\Delta_{14}\right]\neq 0\label{ddd1}\\
   &\mathrm{or} &\fr 1 2 \left[u(h^U \Delta_{41}-s^u_1\Delta_{44})+\om(s^u_1\Delta_{14}-h^U \Delta_{11})+\Delta_{11}\Delta_{44}-\Delta_{41}\Delta_{14}\right]\neq 0.\label{ddd2} \eea It is interesting that the first two terms of (\ref{ddd1}) and (\ref{ddd2}) mean (\ref{ddd}) and (\ref{29dd}), respectively. 

At the one loop level, there must be similar corrections mediated coming from ordinary quarks, exotic quarks as well as both ordinary and exotic quarks in mediations. We must also include general Yukawa couplings connecting flavors, i.e. $h_{ab}\neq 0,\ s_{ab}\neq 0$ for $a\neq b$ to account for the CKM quark mixing matrix as it should be. It is also remarked that the external scalar lines of those diagrams now consist of $\phi$, $\chi$ or both $\chi$ and $\phi$ as well. Totally, we have $48$ diagrams at the one-loop level (24 for up type quark and 24 for down type quark). See Appendix B of \cite{dhhl} for details. Here, for a convenience let us list all those corrections in terms of the relevant matrix elements as given in Appendix \ref{app2}. All the one-loop corrections are taken into account to yield (\ref{upcor1}) explicitly
\bea \Delta_{11} &=& \frac{h^u_{\alpha 1}}{s^U_\alpha}\sum_{i=1}^4 \Delta^i_{14}+\frac{s^u_ 1}{h^U}\sum_{j=5}^8\Delta^j_{14}, \hs\hs\hs
\Delta_{14}= \sum_{k=1}^8 \Delta^k_{14}, \\
  \Delta_{41} &=&\frac{h^u_{\alpha 1}}{s^U_\alpha}\sum_{i=1}^4 \Delta^i_{44}+\frac{s^u_ 1}{h^U}\sum_{j=5}^8\Delta^j_{44}, \hs\hs\hs 
  \Delta_{44} = \sum_{k=1}^8 \Delta^k_{44}.\label{m11}\eea It is easily checked that (\ref{29dd}) is satisfied since \be \frac{h^u_{\alpha 1}}{s^U_\alpha} \neq \frac{s^u_ 1}{h^U},\hs\hs h^u_{\al 1}\neq 0,\hs\hs s^U_{\al}\neq 0,\label{hungdd}\ee in general. This is due to the contribution of $\phi$ to the massless quarks (in addition to $\chi$) as well like we can already see from the Yukawa couplings $h^u_{\al 1}$ and $s^U_\al$ related to this scalar. The system (\ref{ddd}) is always correct even we can check that it is also applied for the special case with flavor diagonalization as presented in \cite{dhhl,natur}.
  
Finally let us check (ii). The determinant equals to
  \be \mathrm{det}M_{\mathrm{up}}=\fr 1 2 \left(\fr{h^u_{\al 1}}{s^U_{\al}}-\fr{s^u_1}{h^U}\right)\left[h^U\sum^4_{i=1}(u\Delta^i_{44}-\om\Delta^i_{14})+\sum^4_{i=1}\sum^8_{j=5}(\Delta^i_{14}\Delta^j_{44}-\Delta^j_{14}\Delta^i_{44})\right],\ee which is always nonzero due to (\ref{hungdd}). In fact, the last factor $[ \cdots]$ can be explicitly given by
  \bea
 && h^U\left\{[(\omega^2+u^2)\lambda_3+u^2\lambda_4 +v^2\lambda_2][u[ I(M^2_{Q_{\alpha 1,3}}, M^2_{d_{i}}, 
M^2_{\phi_3})+  I(M^2_{Q_{\alpha 1,3}}, M^2_{D_{\alpha}}, M^2_{\phi_3})]\right.\crn  && -\left.\omega [ I(M^2_{Q_{\alpha 1,3}}, M^2_{d_{i}}, M^2_{\phi_1}) + I(M^2_{Q_{\alpha 1,3}}, M^2_{D_{\alpha}}, M^2_{\phi_1})] 
 +  M^2_{\chi_3}u[ B(M^2_{Q_1^2}, M^2_{d_i}, M^2_{\chi_2}, M^2_{\phi_3})\right.\crn &&\left.+ B(M^2_{Q_1^2}, M^2_{D_\alpha}, M^2_{\chi_2},M^2_{\phi_3})] - M^2_{\chi_1}\omega[B(M^2_{Q_{\alpha 2}}, M^2_{u_i}, M^2_{\phi^2}, M^2_{\chi_1})+ B(M^2_{Q_{\alpha 2}}, M^2_{U}, M^2_{\phi^2},M^2_{\chi_1})]\right.  \crn
  &&-\left.  (\omega-u)u\omega[A(M^2_{Q_{\alpha1,3}}, M^2_{d_i}, M^2_{\phi_3})+A(M^2_{Q_{\alpha 2}}, M^2_{u_i}, M^2_{\phi_3} )+ M^2_{\phi_1}B(M^2_{Q_{\alpha1,3}}, M^2_{d_i}, M^2_{\phi_3}, M^2_{\phi_1})\right.\crn &&\left.+ M^2_{\phi_1}B(M^2_{Q_{\alpha 2}}, M^2_{u_i}, M^2_{\phi_3}, M^2_{\phi_1})]] \right\}
 +\left\{ u\omega[A(M^2_{Q_{\alpha1,3}}, M^2_{d_i}, M^2_{\phi_3})+A(M^2_{Q_{\alpha 2}}, M^2_{u_i}, M^2_{\phi_3} )\right.  \crn
&&+\left.  M^2_{\phi_1}B(M^2_{Q_{\alpha1,3}}, M^2_{d_i}, M^2_{\phi_3}, M^2_{\phi_1})+ M^2_{\phi_1}B(M^2_{Q_{\alpha 2}}, M^2_{u_i}, M^2_{\phi_3}, M^2_{\phi_1})] \right\}  [[((\omega^2 + u^2)\lambda_1+ v^2 \lambda_3 )\crn
&&\times [ ( I(M^2_{Q_1^{1,3}}, M^2_U, M^2_{\chi_1}) + I(M^2_{Q_1^{1,3}}, M^2_{u_i}, M^2_{\chi_1})) - ( I(M^2_{Q_1^{1,3}}, M^2_U, M^2_{\chi_3}) + I(M^2_{Q_1^{1,3}}, M^2_{u_i}, M^2_{\chi_3}))]] \crn
&&+u[A(M^2_{Q_{\alpha 2}}, M^2_{u_i}, M^2_{\phi^2}) + A(M^2_{Q_{\alpha 2}}, M^2_{U}, M^2_{\phi^2})+  M^2_{\chi_1}[B(M^2_{Q_{\alpha 2}}, M^2_{u_i}, M^2_{\phi^2}, M^2_{\chi_1})\crn
&&+ B(M^2_{Q_{\alpha 2}}, M^2_{U}, M^2_{\phi^2},M^2_{\chi_1})]]- \omega [A(M^2_{Q_1^2}, M^2_{d_i}, M^2_{\chi_2}) + A(M^2_{Q_1^2}, M^2_{D_\alpha}, M^2_{\chi_2})]\crn
&&+ M^2_{\phi_3}[B(M^2_{Q_1^2}, M^2_{d_i}, M^2_{\chi_2}, M^2_{\phi_3})+ B(M^2_{Q_1^2}, M^2_{D_\alpha}, M^2_{\chi_2},M^2_{\phi_3})]]]+\left\{ u \omega[A(M^2_U, M^2_U, M^2_{\chi_3})\right.\crn
&&\left.+A(M^2_U, M^2_{u_i}, M^2_{\chi_3})+  M^2_{\chi_1} B(M^2_U, M^2_U, M^2_{\chi_3}, M^2_{\chi_1})+ M^2_{\chi_1}B(M^2_U, M^2_{u_i}, M^2_{\chi_3}, M^2_{\chi_1})]\right\}  \crn
 &&\times[ [(\omega^2 + u^2)\lambda_3+u^2\lambda_4+v ^2\lambda_2] [( I(M^2_{Q_{\alpha 1,3}}, M^2_{d_{i}}, M^2_{\phi_3}) + I(M^2_{Q_{\alpha 1,3}}, M^2_{D_{\alpha}}, M^2_{\phi_3})) \crn
 &&-( I(M^2_{Q_{\alpha 1,3}}, M^2_{d_{i}}, M^2_{\phi_1}) + I(M^2_{Q_{\alpha 1,3}}, M^2_{D_{\alpha}}, M^2_{\phi_1}))]+\omega[ A(M^2_{Q_{\alpha 2}}, M^2_{u_i}, M^2_{\phi^2}) + A(M^2_{Q_{\alpha 2}}, M^2_{U}, M^2_{\phi^2}) \crn
 &&+ M^2_{\chi_3}[B(M^2_{Q_{\alpha 2}}, M^2_{u_i}, M^2_{\phi^2}, M^2_{\chi_3})+ B(M^2_{Q_{\alpha 2}}, M^2_{U}, M^2_{\phi^2},M^2_{\chi_3})] ] - u [A(M^2_{Q_{\alpha 2}}, M^2_{u_i}, M^2_{\phi^2}) \crn &&+ A(M^2_{Q_{\alpha 2}}, M^2_{U}, M^2_{\phi^2}) + M^2_{\chi_1}[B(M^2_{Q_{\alpha 2}}, M^2_{u_i}, M^2_{\phi^2}, M^2_{\chi_1})+ B(M^2_{Q_{\alpha 2}}, M^2_{U}, M^2_{\phi^2},M^2_{\chi_1})]]] \crn
 &&+ \{ [(\omega^2 + u^2)\lambda_3+u^2\lambda_4+v ^2\lambda_2][ I(M^2_{Q_{\alpha 1,3}}, M^2_{d_{i}}, M^2_{\phi_1}) + I(M^2_{Q_{\alpha 1,3}}, M^2_{D_{\alpha}}, M^2_{\phi_1})] \crn
 &&+  u \left[A(M^2_{Q_{\alpha 2}}, M^2_{u_i}, M^2_{\phi^2}) + A(M^2_{Q_{\alpha 2}}, M^2_{U}, M^2_{\phi^2})+ M^2_{\chi_1}[B(M^2_{Q_{\alpha 2}}, M^2_{u_i}, M^2_{\phi^2}, M^2_{\chi_1})\right.\crn
 &&\left.+ B(M^2_{Q_{\alpha 2}}, M^2_{U}, M^2_{\phi^2},M^2_{\chi_1})]\right] \}  \{[(\omega^2 + u^2)\lambda_1+ v^2 \lambda_3] [ I(M^2_{Q_1^{1,3}}, M^2_U, M^2_{\chi_3}) + I(M^2_{Q_1^{1,3}}, M^2_{u_i}, M^2_{\chi_3})] \crn
 &&+  \omega \left[A(M^2_{Q_1^2}, M^2_{d_i}, M^2_{\chi_2}) + A(M^2_{Q_1^2}, M^2_{D_\alpha}, M^2_{\chi_2})+ M^2_{\phi_3}[B(M^2_{Q_1^2}, M^2_{d_i}, M^2_{\chi_2}, M^2_{\phi_3})\right.\crn &&\left.+ B(M^2_{Q_1^2}, M^2_{D_\alpha}, M^2_{\chi_2},M^2_{\phi_3})]\right]  \} - \{ [(\omega^2 + u^2)\lambda_1+ v^2 \lambda_3] [ I(M^2_{Q_1^{1,3}}, M^2_U, M^2_{\chi_1}) + I(M^2_{Q_1^{1,3}}, M^2_{u_i}, M^2_{\chi_1})] \crn
 &&+  u \left[A(M^2_{Q_1^2}, M^2_{d_i}, M^2_{\chi_2}) + A(M^2_{Q_1^2}, M^2_{D_\alpha}, M^2_{\chi_2})+ M^2_{\phi_1}[B(M^2_{Q_1^2}, M^2_{d_i}, M^2_{\chi_2}, M^2_{\phi_1})\right.\crn
 &&\left.+ B(M^2_{Q_1^2}, M^2_{D_\alpha}, M^2_{\chi_2},M^2_{\phi_1})]\right] \} \{ [(\omega^2 + u^2)\lambda_3+u^2\lambda_4+v ^2\lambda_2][ I(M^2_{Q_{\alpha 1,3}}, M^2_{d_{i}}, M^2_{\phi_1}) \crn
 &&+ I(M^2_{Q_{\alpha 1,3}}, M^2_{D_{\alpha}}, M^2_{\phi_1})] +  u \left[A(M^2_{Q_{\alpha 2}}, M^2_{u_i}, M^2_{\phi^2}) + A(M^2_{Q_{\alpha 2}}, M^2_{U}, M^2_{\phi^2})\right. \crn
 &&+\left. M^2_{\chi_1}[B(M^2_{Q_{\alpha 2}}, M^2_{u_i}, M^2_{\phi^2}, M^2_{\chi_1})+ B(M^2_{Q_{\alpha 2}}, M^2_{U}, M^2_{\phi^2},M^2_{\chi_1})]\right] \},
\eea where the functions $I$, $A$ and $B$ are defined in Appendix \ref{app1}. We conclude that all the quarks in this model can get nonzero masses at the one-loop level. Although the tree level vanishing masses of quarks is not a consequence of the $U(1)_H$ symmetry, this Peccei-Quinn like symmetry is collectively broken at the one-loop level when the quarks get masses.

\subsection{Effective mass operators}

As previous section, the $U(1)_H$ symmetry is spontaneously broken via the collective effects at the one-loop level when all the quarks get mass, i.e. $2 H_\chi + H_\phi \neq 0$. In this section, we will show that all the quarks can get mass via effective mass operators there the $U(1)_H$ breaking is explicitly recognized.  In other words, we will consider effective interactions responsible for fermion masses up to five dimensions. The most general interactions up to five dimensions that lead to fermion masses have the form:
\be \mathcal{L}_Y + \mathcal{L}'_{Y},\label{dong11}\ee
where $\mathcal{L}_Y$ is defined in (\ref{y2}) and $\mathcal{L}'_{Y}$ (five-dimensional effective mass operators) is given by 
\bea  \mathcal{L}'_{Y} &=& \fr{1}{\La}(\overline{Q}_{1L} \phi^*\chi^*)(s'^U U_R + h'^u_a u_{aR})\crn
&& + \fr{1}{\La}(\overline{Q}_{\al L}\phi \chi) (s'^D_{\al \beta} D_{\beta R}+ h'^d_{\al a} d_{aR})\crn
&& + \fr{1}{\La}s'^\nu_{ab}(\overline{\psi}^c_{aL} \psi_{bL} ) (\chi\chi)^*\crn
&& + H.c.\eea Here, as usual we denote $h$ for $\mathcal{L}$-charge conservation couplings and $s$ for violating ones. $\La$ is the cutoff scale which can be taken in the same order as $\om$. It is noteworthy that all the above interactions (as given in $\mathcal{L}'_Y$) are not invariant under $U(1)_H$ since they carry $U(1)_H$ charge proportional to $2H_\chi + H_\phi \neq 0$ like (\ref{ano}). For example, the first interaction has $U(1)_H$ charge: $-H_{Q_1}-H_\phi-H_\chi + H_u=-(2 H_\chi+ H_\phi) $, with the help of eqs (\ref{e1}-\ref{e5}). All those interactions contain $\phi \chi$ combination. Therefore, the fermion masses are generated if both scalars develop  VEV. In this case, the Peccei-Quinn like symmetry $U(1)_H$ is spontaneously broken too. 

Substituting VEVs (\ref{vevp}) into (\ref{dong11}), the mass Lagrangian reads 
\bea \mathcal{L}^{mass}_{fermion}&=& - (\overline{u}_{1L}\ \overline{u}_{2L}\ \overline{u}_{3L}\ \overline{U}_L)M_u (u_{1R}\ u_{2R}\ u_{3R}\ U_R)^T \crn
&&- (\overline{d}_{1L}\ \overline{d}_{2L}\ \overline{d}_{3L}\ \overline{D}_{2L}\ \overline{D}_{3L})M_d (d_{1R}\ d_{2R}\ d_{3R}\ D_{2R}\ D_{3R})^T\crn
&&-\fr 1 2 (\overline{\nu}^c_L\ \overline{N}_R) M_\nu (\nu_L\ N^c_R)^T \crn
&&+H.c.\eea Here the mass matrices of up type quarks $(u_1\ u_2\ u_3\ U)$, down type quarks $(d_1\ d_2\ d_3\ D_2\ D_3)$ are respectively given by 
\be
M_{\mathrm{u}} = \fr{1}{\sqrt{2}}\left(%
\begin{array}{cccc}
  -s^u_{1} u-\fr{1}{\sqrt{2}}\fr{v\om}{\La} h'^{u}_1 & -s^u_{2}u-\fr{1}{\sqrt{2}}\fr{v\om}{\La} h'^{u}_2  & -s^u_{3}u-\fr{1}{\sqrt{2}}\fr{v\om}{\La} h'^{u}_3 & -h^U u-\fr{1}{\sqrt{2}}\fr{v\om}{\La} s'^{U} \\
 h^u_{2 1} v  & h^u_{2 2} v & h^u_{2 3} v & s^U_{2} v \\
 h^u_{3 1} v  & h^u_{3 2} v & h^u_{3 3} v & s^U_{3} v  \\
  -s^u_{1} \om+\fr{1}{\sqrt{2}}\fr{vu}{\La} h'^{u}_1  & -s^u_{2} \om +\fr{1}{\sqrt{2}}\fr{vu}{\La} h'^{u}_2& -s^u_{3} \om +\fr{1}{\sqrt{2}}\fr{vu}{\La} h'^{u}_3& -h^U \om+\fr{1}{\sqrt{2}}\fr{vu}{\La} s'^{U} \\
\end{array}%
\right), \label{upq} \ee
\be M_{\mathrm{d}} =  \fr{-1}{\sqrt{2}}
\left(%
\begin{array}{ccccc}
  h^d_{1} v  &  h^d_{2} v  &  h^d_{3} v  &  s^D_{2} v  & s^D_{3} v \\
  s^d_{21} u +\fr{1}{\sqrt{2}}\fr{v\om}{\La} h'^{d}_{21} &  s^d_{22} u +\fr{1}{\sqrt{2}}\fr{v\om}{\La} h'^{d}_{22} &  s^d_{23} u +\fr{1}{\sqrt{2}}\fr{v\om}{\La} h'^{d}_{23}&  h^D_{22} u +\fr{1}{\sqrt{2}}\fr{v\om}{\La} s'^{D}_{22}&  h^D_{23} u +\fr{1}{\sqrt{2}}\fr{v\om}{\La} s'^{D}_{23}\\
   s^d_{31} u +\fr{1}{\sqrt{2}}\fr{v\om}{\La} h'^{d}_{31}&  s^d_{32} u +\fr{1}{\sqrt{2}}\fr{v\om}{\La} h'^{d}_{32}&  s^d_{33} u  +\fr{1}{\sqrt{2}}\fr{v\om}{\La} h'^{d}_{33}& h^D_{32} u +\fr{1}{\sqrt{2}}\fr{v\om}{\La} s'^{D}_{32} & h^D_{33} u +\fr{1}{\sqrt{2}}\fr{v\om}{\La} s'^{D}_{33} \\
   s^d_{21} \om-\fr{1}{\sqrt{2}}\fr{vu}{\La} h'^{d}_{21} & s^d_{22} \om -\fr{1}{\sqrt{2}}\fr{vu}{\La} h'^{d}_{22}& s^d_{23} \om -\fr{1}{\sqrt{2}}\fr{vu}{\La} h'^{d}_{23}& h^D_{22} \om -\fr{1}{\sqrt{2}}\fr{vu}{\La} s'^{D}_{22}& h^D_{23} \om -\fr{1}{\sqrt{2}}\fr{vu}{\La} s'^{d}_{23} \\
  s^d_{31} \om -\fr{1}{\sqrt{2}}\fr{vu}{\La} h'^{d}_{31}& s^d_{32} \om -\fr{1}{\sqrt{2}}\fr{vu}{\La} h'^{d}_{32}& s^d_{33} \om -\fr{1}{\sqrt{2}}\fr{vu}{\La} h'^{d}_{33}& h^D_{32} \om -\fr{1}{\sqrt{2}}\fr{vu}{\La} s'^{D}_{32}& h^D_{33}\om-\fr{1}{\sqrt{2}}\fr{vu}{\La} s'^{D}_{33} \\
\end{array}%
\right).
 \label{downq} \ee
And the mass matrix for neutrinos $(\nu_1\ \nu_2\ \nu_3\ N^c_1\ N^c_2\ N^c_3)$ has the form
\be M_\nu = -\left(\begin{array}{cc}s'^\nu \fr{u^2}{\La}& (\fr{u\om}{\La} s'^\nu +\sqrt{2} v h^\nu)^T \\
\fr{u\om}{\La} s'^\nu +\sqrt{2} v h^\nu & s'^\nu \fr{\om^2}{\La}\end{array}\right).\label{lepm}\ee     
Notice that $h^\nu$ is a matrix of flavor indices $h^\nu_{ab}$ and antisymmetric in $a$ and $b$, while $s'^\nu$ is also a matrix of three flavors but symmetric in these indices.  

Three remarks are in order
\ben
\item Up quarks: If there is no correction, i.e. $s',h'=0$, the mass matrix (\ref{upq}) has the first line and the fourth line in proportion (degeneracy) that means that one up quark is massless, as mentioned \cite{dhhl}. The presence of corrections, i.e. $s',h'\neq 0$, will separate that degeneracy. Indeed, the first and fourth lines are now in proportion only if $s^u_1/h'^u_1 = s^u_2/h'^u_2 = s^u_3/h'^u_3=h^U/s'^U$ which is not the case in general. The up quark type mass matrix is now most general that can be diagonalized to obtain the masses of exotic $U$ and ordinary $u_{1,2,3}$.    
\item Down quarks: The second and the fourth lines as well as the third and the fifth lines have the same status as in the up quark type. All these degeneracies are separated. Consequently we have the most general mass matrix for down quark type. 
\item Neutrino: It is interesting that the neutrinos also get mass via the breaking of Peccei-Quinn like symmetries in association with the lepton number breaking. If the correction $s'^\nu$ vanishes, the neutrinos have one zero mass and two degeneracies which do not coincide with the data \cite{dls2}. The presence of correction implies a seesaw mechanism. The mass matrix of observed neutrinos is given by
\be M_{active}=2\fr{v^2}{\om}\fr{\La}{\om} h^\nu (s'^\nu)^{-1}h^\nu. \ee Here the factor $\La/\om$ is in order of unity. The neutrino masses are small due to the suppression of $v/\om$. The scale $\om,\La$ should be very high so that the model takes a consistency with the low energy phenomenologies.  
\een

Using the $U(1)_H$ violating triple scalar interactions as mentioned above, those effective mass operators with five dimensions can be explicitly understood as derived from two-loop radiative corrections responsible for the quark masses, with the assumption that $U(1)_H$ was broken in the scalar potential first, in similarity to the radiative Majorana neutrino masses via lepton violating triple scalar potentials in Zee-Babu model \cite{zeebabu}. It is noted that the above one-loop corrections can be also translated via the language of effective operators with six dimensions before the $U(1)_H$ breaking happens. A complete calculation of all the corrections presented above as well as obtaining the quark masses and mixing is out of scope of this work. It should be published elsewhere \cite{dlh}.

\section{\label{con}Conclusions}

As any other 3-3-1 models, the economical 3-3-1 model naturally contains an infinity of $U(1)_H$ symmetries like Peccei-Quinn symmetry with just its scalar content, which is unlike the case of the standard model. In contradiction to other extensions of the standard model including ordinary 3-3-1 models, the economical 3-3-1 model has interesting features as follows
\ben
\item There is no residual symmetry of $U(1)_H$ after the scalars getting VEVs.
\item There is no axion associated with $U(1)_H$ breaking.
\item There is no Majoron associated with lepton number breaking.
\item The vanishing of quark masses at the tree-level is not a resultant from $U(1)_H$. It is already a consequence of the minimal scalar content under the model gauge symmetry.   
\item All the quarks can get nonzero masses at the one-loop level, there the $U(1)_H$ symmetry is obviously broken. 
\item The mass generation for neutrinos also breaks $U(1)_H$.   
\een  

By this work, it is to emphasis that the economical 3-3-1 model can work with only two scalar triplets. All the fermions can get consistent masses \cite{dhhl,dls2}. A further analysis can show observed flavor mixings as indicated by the CKM matrix and PMNS matrix. Also, with the minimal scalar sector the model is very predictive which is worth to be searched for at the current colliders \cite{dlh}.  

With the above conclusions, it is to emphasis that the statements in \cite{natur} such as unique solution of $U(1)_H$, existence of a residual symmetry of $U(1)_H$, the masslessness of quarks due to that supposed residual symmetry, and one loop corrections to up type quark sector are incorrect or incomplete. The addition of scalars to the economical 3-3-1 model as given in \cite{natur} is dynamically not required since this model as seen can work well by itself.     

\section*{Acknowledgement}

PVD and HNL would like to thank Professors Hsiang-Nan Li, Hai-Yang Cheng, Hoi-Lai Yu, and Tzu-Chiang Yuan for their warm hospitality and financial support during the stay where this work was initiated. We would also like to thank Members of the high energy physics group, specially Vu Thi Ngoc Huyen, at Institute of Physics, VAST for useful discussions. This work was supported by National Foundation for Science and Technology Development of Vietnam (NAFOSTED) under Grant No. 103.03-2011.35.

\appendix

\section{\label{app1}Integrations}

The functions $A(a,b,c)$, $B(a,b,c,d)$ and $I(a,b,c)$ as appeared in the text are given by
\bea 
 A(a,b,c)&\equiv& \int\fr{d^4p}{(2\pi)^4}\fr{1}{(p^2-a)(p^2-b)(p^2-c)}\crn
 &=&\fr{-i}{16\pi^2}\left\{\fr{a\ln
a}{(a-b)(a-c)}+\fr{b\ln
b}{(b-a)(b-c)}+\fr{c\ln c}{(c-b)(c-a)}\right\},\label{eqn7031}\\
 B(a,b,c,d)&\equiv& \int\fr{d^4p}{(2\pi)^4}\fr{1}{(p^2-a)(p^2-b)(p^2-c)(p^2-d)}\crn
 &=&\fr{-i}{16\pi^2}\left\{\fr{a\ln
a}{(a-b)(a-c)(c-d)}+\fr{b\ln
b}{(b-a)(b-c)(b-d)}\right. \crn && \left.+\fr{c\ln c}{(c-b)(c-a)(c-d)}+ \fr{d\ln d}{(d-b)(d-a)(d-c)}\right\},\label{eqn7031}\\
I(a,b,c)&\equiv&
\int\fr{d^4p}{(2\pi)^4}\fr{p^2}{(p^2-a)^2(p^2-b)(p^2-c)}\crn
&=&
\fr{-i}{16\pi^2}\left\{\fr{a(2\ln a +1)}{(a-b)(a-c)} -\fr{a^2(2a -b
-c)\ln a }{(a-b)^2(a-c)^2} + \fr{b^2\ln
b}{(b-a)^2(b-c)}\right.\crn 
&&\left.+\fr{c^2\ln
c}{(c-a)^2(c-b)}\right\}.\label{eqn7033}\eea

\section{\label{app2}Corrections}

The one-loop corrections to the mass matrix $M_{uU}$ are presented by the diagrams as follows:

\begin{figure}[htbp]
\includegraphics{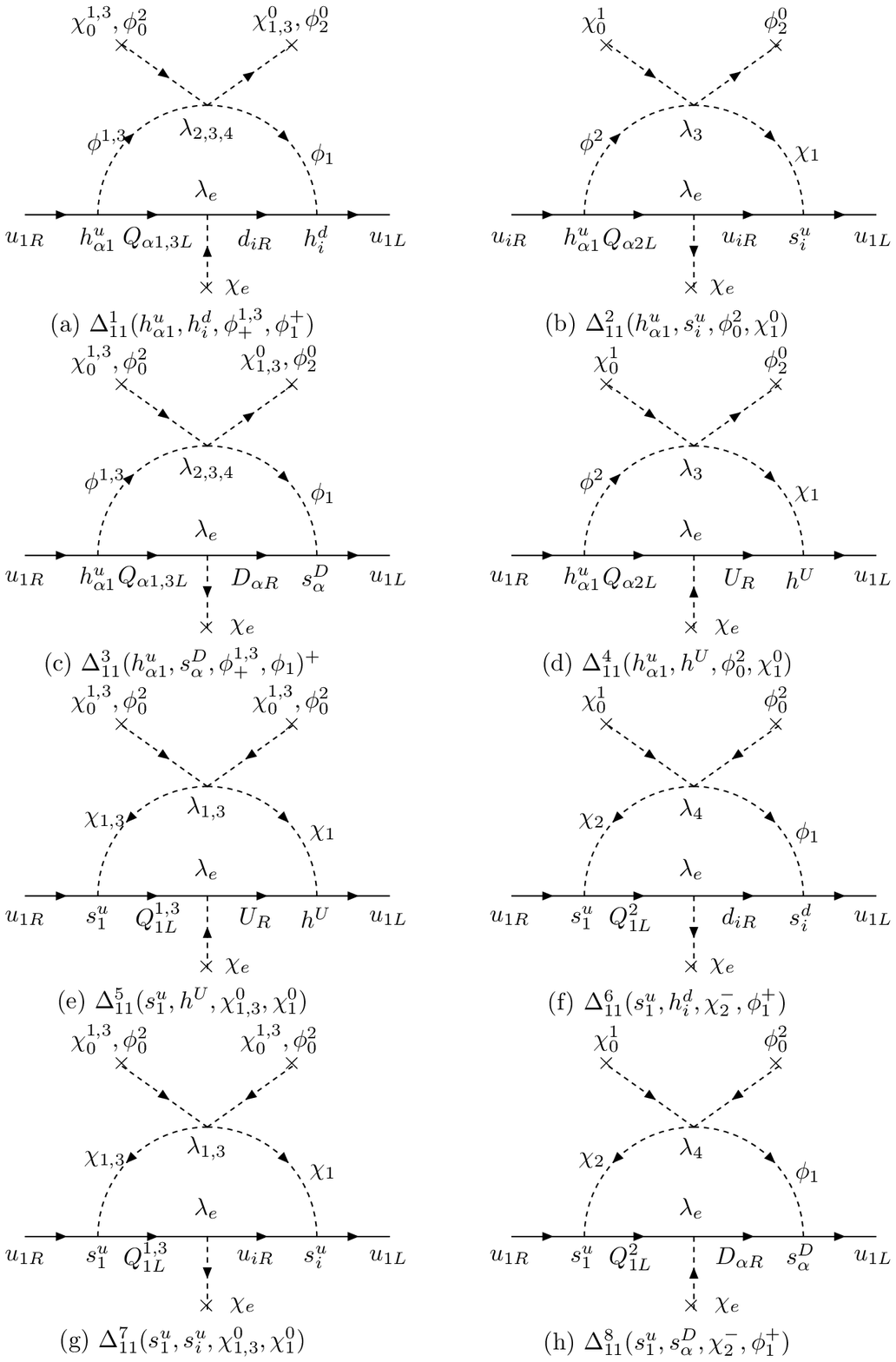}
 \caption[]{\label{figh4} Corrections to $(M_{uU})_{11}$}
\end{figure}

\begin{figure}[htbp]
\includegraphics{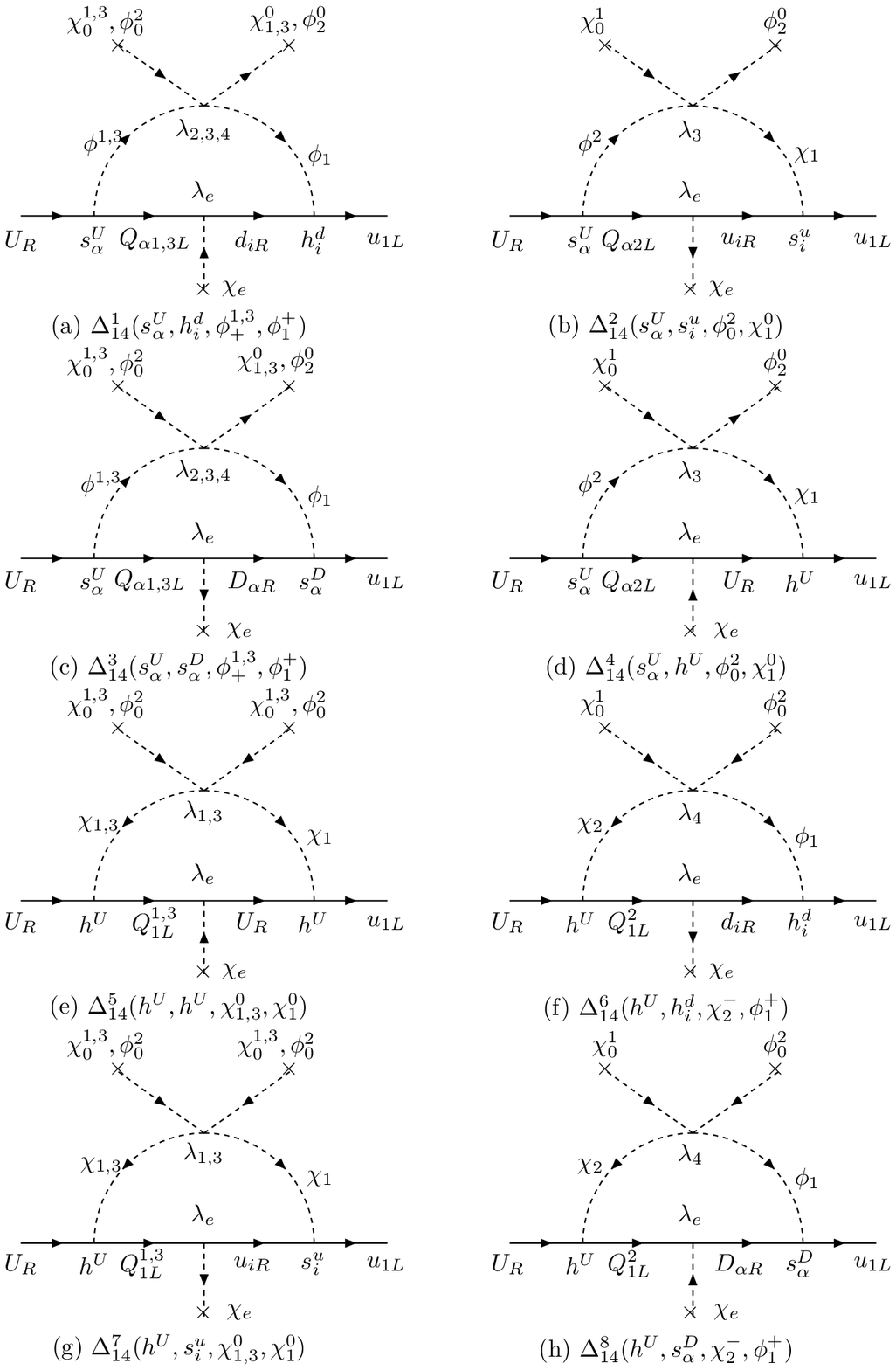}
\caption[]{Corrections to $(M_{uU})_{12}$.}
\end{figure}

\begin{figure}[htbp]
\includegraphics{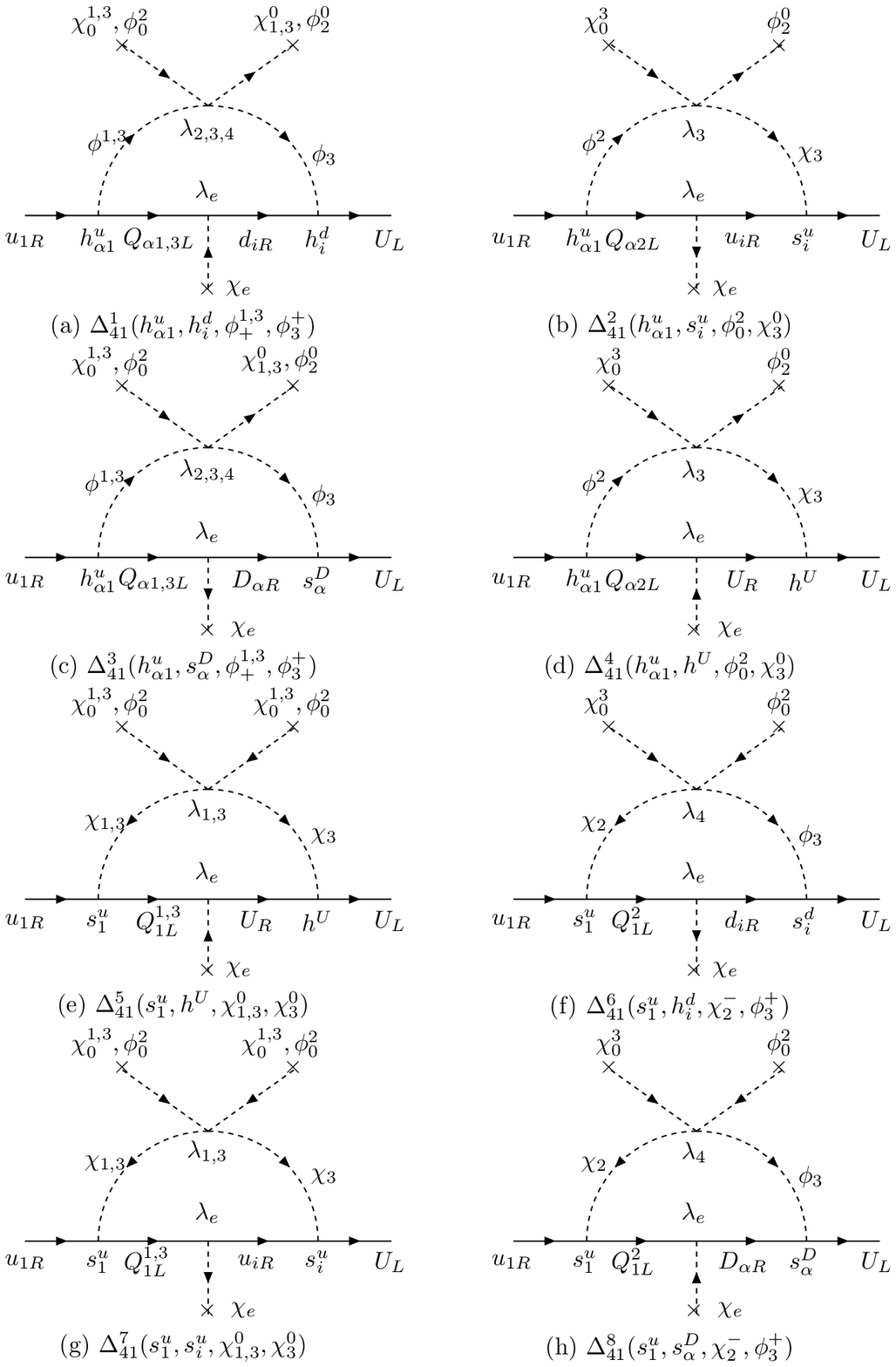}
\caption[]{Corrections to $(M_{uU})_{21}$.}
\end{figure}

\begin{figure}[htbp]
\includegraphics{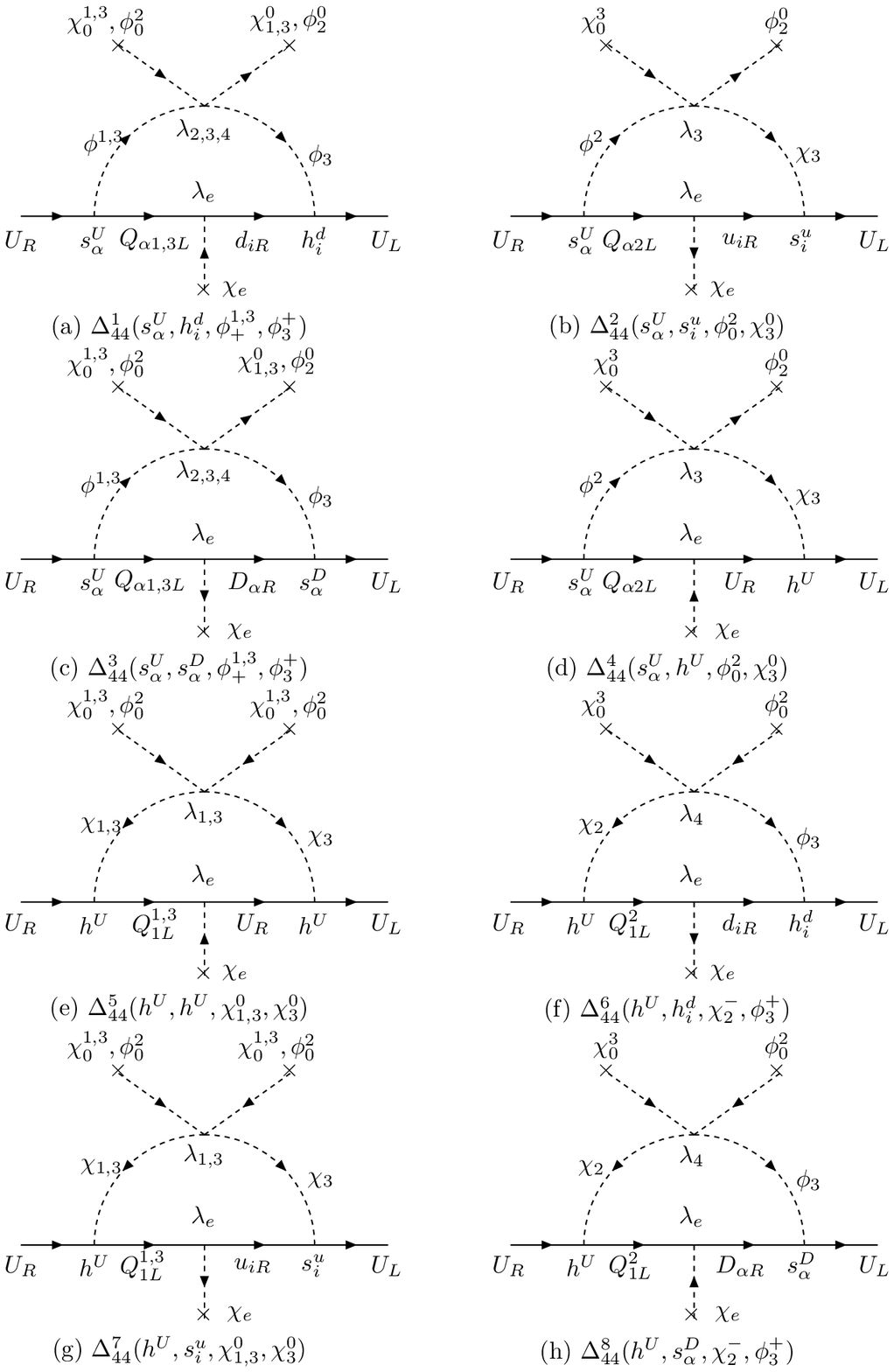}
\caption[]{Corrections to $(M_{uU})_{22}$.}
\end{figure}

\end{document}